\documentclass[10pt,twocolumn,showpacs,preprintnumbers,amsmath,amssymb]{revtex4}
\usepackage{url}
\usepackage{stmaryrd}
\usepackage{graphicx}
\usepackage{setspace}
\usepackage{amsthm}
\usepackage{enumerate}
\newtheorem{theorem}{Theorem}[section]
\newtheorem{lemma}[]{Lemma}
\newtheorem{proposition}[]{Proposition}
\newtheorem{corollary}[theorem]{Corollary}
\newtheorem{definition}[]{Definition}
\newtheorem{remark}[theorem]{Remark}

\newcommand*{\textfrac}[2]{{#1}/{#2}}

\newcommand*{\cE}{\mathcal{E}}
\newcommand*{\cF}{\mathcal{F}}

\newcommand*{\cQ}{\mathcal{Q}}

\newcommand*{\cS}{\mathcal{S}}
\newcommand*{\cU}{\mathcal{U}}

\newcommand*{\cV}{\mathcal{V}}
\newcommand*{\cW}{\mathcal{W}}
\newcommand*{\cX}{\mathcal{X}}
\newcommand*{\cY}{\mathcal{Y}}
\newcommand*{\cZ}{\mathcal{Z}}

\newcommand*{\tr}{\mathsf{tr}}
\newcommand*{\sbin}{\{0,1\}}
\newcommand*{\ket}[1]{|#1\rangle}
\newcommand*{\bra}[1]{\langle #1|}
\newcommand*{\proj}[1]{\ket{#1}\bra{#1}}

\newcommand*{\id}{\mathsf{1}}

\newcommand*{\ExpE}[1]{\underset{#1}{\mathbb{E}}}
\newcommand*{\uniform}{\cU}
\newcommand*{\Ext}{\mathsf{e}}
\newcommand*{\bigExt}{\mathsf{E}}
\newcommand*{\nonun}[3]{d({#2\shortleftarrow #3})}
\newcommand*{\nonuncond}[4]{d({#2\shortleftarrow #3}|#4)}
\newcommand*{\hguesscond}[3]{H_{\mathsf{g}}(#1\shortleftarrow #2|#3)} 
\newcommand*{\hguessn}[2]{H_{\mathsf{g}}(#1\shortleftarrow #2)} 
\begin{document}

\title{
The Bounded Storage Model in The Presence of a Quantum Adversary
}

\author{Robert K\"onig$^1$ and Barbara M. Terhal$^2$}
\affiliation{\vspace*{1.2ex}
            \hspace*{0.5ex}{$^1$ Centre for Quantum Computation, DAMTP, University of Cambridge, Cambridge CB3 0WA, UK}}
\affiliation{\vspace*{1.2ex}
            \hspace*{0.5ex}{$^2$IBM Watson Research Center,
P.O. Box 218, Yorktown Heights, NY 10598, USA}}
\date{\today}

\begin{abstract}
An extractor is a function $\bigExt$ that is used to extract randomness. Given an imperfect random source $X$ and a uniform seed $Y$, the output $\bigExt(X,Y)$ is close to uniform. We study properties of such functions in the presence of prior quantum information about $X$, with a particular focus on cryptographic applications. We prove that certain extractors  are  suitable for key expansion in the bounded storage model where the adversary has a limited amount of quantum memory. For extractors with one-bit output we show that the extracted bit is essentially equally secure as in the case where the adversary has classical resources. We prove the security of certain constructions that output multiple bits in the bounded storage model.
\end{abstract}

 \maketitle

\section{Introduction}
The aim of {\em randomness extraction} is to generate  ``almost uniform'' randomness given an imperfect source of randomness $X$. The term ``extractor'' 
is generally used to describe a procedure which accomplishes this task; more formally, an extractor is a (deterministic) function $\bigExt:\cX\times\cY\rightarrow \cZ$ which, when applied to an imperfect source $X$ and a uniform and independent seed $Y$, yields an output $Z:=\bigExt(X,Y)$ which is close to being uniformly distributed on $\cZ$.  Such an extractor is characterized by a number of parameters. Among these are the amount of randomness $Y$ that is required, the amount  of randomness $Z$ produced, and, most importantly,  the character of the sources $X$ which lead to almost uniform output.  A very general class of sources are the {\em weak} sources $X$, characterized by a lower bound on the {\em min-entropy} $H_{\infty}(X):=-\log \max_x P_X(x)$. Correspondingly, a $(k,\varepsilon)$-extractor~\cite{NisZuc96}  commonly refers to an extractor which, for any input distribution $P_X$ with $H_\infty(X)\geq k$, outputs $\varepsilon$-uniform randomness $Z$. 

Besides purifying randomness, extractors are an essential tool in computer science, in particular in complexity theory and cryptography. Correspondingly, the study of such extractors has been a major research topic in recent years, and much understanding has been gained (see~\cite{Shal02} for a review). For applications in computer science, the  challenge is to find explicit, efficiently computable extractors with good parameters.

In a cryptographic context, a certain variant of the concept of a $(k,\varepsilon)$-extractor is of particular importance. These are called {\em strong} extractors; they have the additional property that even the pair $(Y,\bigExt(X,Y))$ is $\varepsilon$-close to uniform. This means 
for example that $(Y,\bigExt(X,Y))$ can be used to encrypt a message $M=(M_1,M_2)$ using a one-time pad~\cite{Vernam26} as $C=(C_1,C_2)=(M_1\oplus Y,M_2\oplus \bigExt(X,Y))$. An adversary who learns the cipher-text $C$ as well as  the message $M_1$  (and thus the seed $Y$) will be completely ignorant of the content of the remaining message $M_2$. Expressed differently, the pair $(Y,\bigExt(X,Y))$ is a key with  {\em universally composable} security~\cite{PfiWai00a,Canetti00}.

A  more striking application of strong extractors in cryptography is privacy amplification, introduced by Bennett, Brassard and Robert~\cite{BeBrRo88} and further analyzed in~\cite{BBCM95}. This refers to a technique that allows  two parties, Alice and Bob, to generate a secret key  $Z$ from a shared random variable $X$ about which the adversary has partial information $E$. The only assumption is that the parties are connected by an authentic but otherwise completely insecure channel. The key $Z$ is then obtained as follows: Alice generates an independent uniform seed $Y$ and sends it over the channel. Subsequently, both parties apply a strong extractor to get $Z:=\bigExt(X,Y)$. The security of $Z$ when used as a secret key directly follows from the properties of the strong extractor, assuming a certain bound on the information $E$ of the adversary.

Apparently related to privacy amplification, but conceptually quite different, is Maurer's bounded storage model~\cite{Maurer92b}. The first security proof for general adversaries in this model was obtained by Aumann, Ding and Rabin~\cite{ADR:bs} and essentially optimal constructions were subsequently found in a sequence of papers~\cite{DziMau02,Lu02,Vadhan03}. Its aim is not {\em key extraction}, but {\em key expansion}.  In this setting a large amount of randomness $X$ is publicly, but only temporarily available. Alice and Bob use a previously shared (short) secret key $Y$ to obtain additional key bits $Z=\bigExt(X,Y)$ using a strong extractor. The seed $Y$ remains hidden to the adversary until (possibly) after the execution of the protocol. The adversary is assumed to have only a bounded amount of storage (which may be much larger than the honest parties' memory). As a result, his information $E$ about $X$ is limited, once $X$ becomes inaccessible, and by the properties of the extractor, $Z$ can be shown to be secure even if he later obtains the seed $Y$ (this was referred to as ``everlasting security'' in~\cite{ADR:bs}).

From a cryptographic viewpoint, a natural generalization of these
scenarios is arrived at by allowing the adversary to have quantum information $Q$ instead of only classical information $E$ about $X$. This modification is not merely of theoretical interest. Indeed, the only construction proved to be secure~\cite{BenOr02,KoMaRe05,RenKoe05,Renner05} for privacy amplification has found various applications in quantum cryptography. Besides simplifying and improving security proofs for quantum key distribution~\cite{ChReEk04,Renner05},
the quantum version of privacy amplification has been used to derive both possibility~\cite{DFSS05} and impossibility~\cite{BCHLW05} results for tasks such as bit commitment or oblivious transfer.

While the problem of constructing strong extractors is well-studied,  little is known about the security resulting from their use in a quantum context. For the bounded storage model, Gavinsky, Kempe and de Wolf~\cite{GKW:sepcc} recently gave an example of an extractor which yields a classically secure key, but is completely insecure against an adversary with a similar amount of quantum storage. There is no construction for the bounded storage model that is known to be secure against a quantum adversary.

In this paper, we study  properties of strong extractors in a context where the adversary has quantum information, with the two cryptographic settings described in mind. We give the first constructions of extractors that are usable in the bounded storage model against a quantum adversary, and we show that certain strong extractors generate secure key bits in the setting of privacy amplification. This allows to reduce the amount of communication needed in certain applications. Our constructions achieve the most desirable type of security, that is, the extracted keys are {\em universally composable}~\cite{BHLMO05,RenKoe05,KRBM05}.

\subsubsection*{Outline}
In Section~\ref{sec:extrunv} we introduce the relevant definitions. In Section~\ref{sec:binary} we show that any strong extractor which outputs a single bit yields essentially the same degree of security in a cryptographic setting,
irrespective of whether the adversary has quantum or classical information. We then use a hybrid argument in Section~\ref{sec:nonbinary} to obtain extractors that output several bits. In Section~\ref{sec:boundedstorage} we explain how these extractors can be used in the bounded storage model. Finally, we show that general strong extractors can be used in the setting of privacy amplification in Section~\ref{sec:tomography}. We conclude in Section~\ref{sec:conclusions}.

\subsection{Notation}
Throughout this paper, all logarithms are binary, i.e., to base $2$.
For a random variable $X$ with range~$\cX$, we define the {\em min-entropy} of $X$ as
$H_\infty(X):=-\log\max_x P_X(x)$. More generally, for a quantum
state $\rho_Q$ on a Hilbert space $\cQ$, $H_\infty(Q)$ is the min-entropy of the distribution
of eigenvalues of $\rho_Q$. Analogously, the max-entropy is defined
as $H_0(X):=\log |\mathrm{supp}(P_X)|=\log |\cX|$ and $H_0(Q):=\log\mathrm{rank}(\rho_Q)$, respectively. Expressed differently, $H_0(Q)$ is the number of qubits constituting system~$Q$. For a function $g:\cX\rightarrow\mathbb{R}$, we denote by 
\[
\ExpE{x\leftarrow P_X} [g(x)]:=\sum_{x\in\cX}P_X(x)g(x)\ 
\]
the expectation of $g(X)$ over a random choice of  $x\leftarrow P_X$. We also use the notation $P_X\cdot P_Y$ to refer to the joint distribution of two independent random variables $X$ and $Y$, that is, $\Pr[X=x,Y=y]=P_X(x)\cdot P_Y(y)$ for all $(x,y)\in\cX\times\cY$.

In the sequel, $Q$ refers to a quantum system, whereas $E$, $V$,
$W$, $X$, $Y$ and $Z$ are assumed to be classical. Slightly abusing
notation, we sometimes refer to the Hilbert space corresponding to a
classical-quantum state (cq-state) $\rho_{XQ}$ by $\cX\otimes\cQ$. We denote the
completely mixed state on $\cX$ by $\rho_{\cU_{\cX}}$.

We will sometimes use classical-quantum states  with multipartite
classical parts, e.g., a ccq-state $\rho_{XYQ}$. For such a state
$\rho_{XYQ}$, we say that  $Y\leftrightarrow X\leftrightarrow Q$
forms a Markov chain if it has the form
\begin{align}\label{eq:markovchainstate}
\rho_{XYQ}=\sum_{x,y} P_{XY}(x,y)\proj{xy}\otimes \rho_x\
\end{align}
for some states $\{\rho_x\}_{x\in\cX}$ on $\cQ$.  A state with this property  defines a distribution $P_{XY}$, which defines the conditional distributions $P_{X|Y=y}$ and, for any function $f:\cX\times\cY\rightarrow \cZ$, the distribution $P_{f(X,Y)XY}$. The corresponding conditional states $\rho_{XQ|Y=y}$ are obtained by making the approriate replacement in Eq.~\eqref{eq:markovchainstate}, i.e.,
\[
\rho_{XQ|Y=y}=\sum_{x} P_{X|Y=y}(x)\proj{x}\otimes\rho_x\ .
\]
Similarly, we can define the cccq-state
\[
\rho_{f(X,Y)XYQ}=\sum_{x,y} P_{XY}(x,y)\proj{f(x,y)xy}\otimes \rho_x\ ,
\]
which in turn gives rise to states such as $\rho_{f(X,y)XQ|Y=y}$.

We will use the trace norm $\|A\|:=\frac{1}{2}\tr(\sqrt{A^\dagger
A})$ for any operator $A$. Note that if $\rho_{XQ}$ and
$\sigma_{X'Q'}$ are cq-states on $\cX\otimes\cQ$, then
\begin{align}\label{eq:rhosigmadistance}
\|\rho_{XQ}-\sigma_{X'Q'}\|=\sum_{x\in\cX} \|P_X(x)\rho_x-P_{X'}(x)\sigma_x\|\ .
\end{align}
For two probability distributions $P$ and $Q$ on $\cX$,
the trace norm of their difference (when identifying the distribution with a state), i.e.
$\|P-Q\|:=\frac{1}{2}\sum_{x\in\cX} |P(x)-Q(x)|$ is also known as
the variational distance.

Let $\rho_{XQ}=\sum_{x\in\cX} P_X(x)\proj{x}\otimes\rho_x$ be a cq-state. Consider a fixed POVM $\cE:=\{E_z\}_{z\in\cZ}$ on $\cQ$. We denote by $P_{XZ}\equiv\rho_{X\cE(Q)}$ the joint distribution of $X$ and the measurement outcome, i.e.,
\[
P_{Z|X=x}(z)=\tr(E_z\rho_x)
\]
for every $z\in\cZ$ and $x\in\cX$.

We will often encounter scalar quantities $d$ that are functions of a given distribution or a quantum state, i.e., $d=d(P_X)$ or $d=d(\rho_Q)$. In these cases, we use the shorthand $d(X)$ or $d(Q)$. Similarly, we write $d(Q|W=w)$ instead of $d(\rho_{Q|W=w})$. More generally, we will consider quantities that depend on a specific bipartition of a state $\rho_{ZE}$ into $Z$ and $E$; in these cases, we write
 $\nonun{P}{Z}{E}$. Again, we use the notation  $\nonuncond{P}{Z}{E}{W=w}$ to denote the corresponding quantity for the conditional state $\rho_{ZE|W=w}$.

\section{Extractors and Secret Keys\label{sec:extrunv}}
\label{sec:propextract}

\subsection{Classical adversaries}

Before reviewing the definition of strong extractors and various of
their basic properties, let us introduce a short-hand notation for
the {\em non-uniformity}, a quantity which measures the extent to
which a probability distribution of a random variable
$Z$ deviates from the uniform distribution, possibly given another
random variable $E$:

\begin{definition}\label{def:classicalextractor}
Let $P_{ZE}$ be an arbitrary distribution. The {\em non-uniformity
$\nonun{P}{Z}{E}$ of $Z$ given $E$} is defined as
\[
\nonun{P}{Z}{E}:=\|P_{ZE}-P_{\uniform_\cZ}\cdot P_E\|.
\]
Here $P_E$ is the marginal distribution of $P_{ZE}$, and $P_{\uniform_\cZ}$ denotes the uniform distribution on $\cZ$.
\end{definition}
Note that $d(Z)$ is simply the distance of the distribution $P_Z$ from the uniform distribution. A strong extractor can then be defined as follows.
\begin{definition}
A {\em strong $(k,\varepsilon)$-extractor} is a function
$\bigExt:\cX\times\cY\rightarrow \cZ$ with the property that 
\begin{align}\label{eq:distancetoidealcla}
\nonun{\rho}{\bigExt(X,Y)}{Y}= \|P_{\bigExt(X,Y)Y}-P_{\uniform_{\cZ}}\cdot
P_{\uniform_{\cY}}\| \leq \varepsilon
\end{align}
for all distributions $P_X$ with $H_\infty(X)\geq k$. Here $Y$ is
independent of $X$ and uniformly distributed on $\cY$.
\end{definition}

The definition implies that $\bigExt(X,y)$ is
close to being uniformly distributed on $\cZ$ on average over the
random choice of $y\leftarrow P_Y$ (cf. Eq.~\eqref{eq:expynonuniformity}). In other words, if $X$ is chosen according
to $P_X$ and $Y$ is uniformly distributed and independent of $X$,
then $\bigExt(X,Y)$ is indistinguishable from uniform, even given $Y$.

In a cryptographic setting, the security of the extracted key
$Z:=\bigExt(X,Y)$ with respect to an adversary who is given $Y$ is
exactly characterized by Eq.~\eqref{eq:distancetoidealcla}. Indeed,
expression~\eqref{eq:distancetoidealcla} quantifies how
distinguishable the real system (consisting of $(Z,Y)$) is from the
ideal system, in which $Z$ is uniformly distributed and independent
of $Y$. This is easily generalized to a setting where the adversary
is given additional information about $X$. The additional
information can be in the form of a classical random variable (i.e., bits) that is jointly distributed with $X$ or a quantum state (i.e., qubits).

In case the adversary has classical information about $X$ expressed by 
a random variable $E$, one can show that this simply
reduces the min-entropy of $X$. If $E$ gives little information about $X$ it follows that even given $E$ and $Y$, the extracted bits look random. This intuition is made explicit in the following proposition (all proofs in this section can be found in Appendix \ref{app:a}):
\begin{proposition}\label{lem:guessingprobability}
Let $\bigExt:\cX\times\cY\rightarrow\cZ$ be a strong
$(k,\varepsilon)$-extractor. Let $P_{XE}$ be a distribution with
\begin{align}\label{eq:maximumsuccprob}
\hguessn{X}{E}\geq k+\log\textfrac{1}{\varepsilon}\ .
\end{align}
Here the guessing-entropy $\hguessn{X}{E}$ of $X$ given $E$ is
defined as
\[
\hguessn{X}{E}:=-\log \max_{\hat{X}}\Pr[X=\hat{X}]\ ,
\]
where the maximum is taken over all random variables $\hat{X}$ such
that $X\leftrightarrow E\leftrightarrow \hat{X}$ forms a Markov
chain. Then 
\[
\nonun{P}{\bigExt(X,Y)}{YE}\leq 2\varepsilon\ ,
\]
where $P_{YXE}:=P_{\uniform_{\cY}}\cdot P_{XE}$.
\end{proposition}
Note that if $E$ is trivial or independent of $X$ the guessing entropy $\hguessn{X}{E}$ of $X$ given $E$ is equal to the min-entropy $H_\infty(X)$ of $X$. 
Proposition~\ref{lem:guessingprobability} can be applied in the bounded storage model because the limitation on the adversary's storage implies that his information about $X$ is bounded. More precisely, the guessing probability has the following intuitive property. Any (additional) piece of
information $W$ does not  increase the success probability in
guessing by a significant amount if the size of $W$ is small. More trivially, independent information $V$ does not affect the guessing probability. We express this formally in Lemma~\ref{lem:wolflemmaclassical}; versions of this statement are implicit in~\cite{NisZuc96}, and more explicitly given in~\cite{Wol99}.
\begin{lemma}\label{lem:wolflemmaclassical}
Consider a distribution $P_{XVWE}$ with
$P_{XV}=P_{X}\cdot P_V$ and $VW\leftrightarrow X\leftrightarrow E$.
Then
\[
\hguessn{X}{VWE}\geq \hguessn{X}{E}-H_0(W)\ .
\]
In particular, for every $\varepsilon\geq 0$,\[
\hguesscond{X}{E}{V=v,W=w}\geq
\hguessn{X}{E}-H_0(W)-\log\textfrac{1}{\varepsilon}
\]
with probability at least $1-\varepsilon$ over $(v,w)\leftarrow
P_{VW}$.
\end{lemma}

\subsection{Quantum adversaries}
Let us now discuss the challenge posed by quantum adversaries. Our aim is to show that, similarly as in the classical case, the extracted bits $\bigExt(X,Y)$ are secure even if the adversary is given $Y$.  Such an adversary prepares a quantum state $\rho_x$ on $\cQ$ that depends on $X=x$. To obtain maximal information about $\bigExt(X,Y)$,  he performs a measurement on his quantum system $Q$ which depends on $Y$. As a result, his (classical) information $E$ is no longer independent of $Y$.
This means that we cannot view this as merely a reduction of the entropy of the
source $X$. Thus we cannot directly prove a statement like
Lemma~\ref{lem:guessingprobability} when $E$ is replaced by a quantum system $Q$. In particular, due to the effect of locking~\cite{divincenzo+:locking},
we know that there exist short classical keys ($Y$) that can unlock
a lot of classical information (about $X$) stored in a quantum
system $Q$. In the first part of this paper we will show that if the
extractor $\bigExt$ extracts a single bit, we can preclude such
locking effect (Theorem~\ref{thm:mainquantumbinary}). 

Before embarking on this analysis, we point out the following straightforward result. {\em If} the adversary's measurement does not depend on $Y$ we can essentially apply the classical security proofs. That is, the adversary's measurement produces
some classical information $E$ which can be viewed as reducing the
entropy of the source $X$. If the size of the quantum system is
sufficiently small, then the random variable $E$ does not give
much information about $X$ and therefore the extracted bits
look random even to such an adversary. These statements are expressed
in the following two lemmas.

Note that we can generalize the
guessing-entropy of $X$ given $Q$ to the case where $Q$ is a quantum system. Let
$\rho_{XQ}:=\sum_{x\in\cX} P_X(x)\proj{x}\otimes\rho_x$ be a
cq-state. Then
\begin{align}\label{eq:hguessQdef}
\hguessn{X}{Q}:=-\log \max_{\cE}\sum_{x\in\cX}
P_X(x)\tr(E_x\rho_x)\ , \end{align} where the maximum is taken
over all POVMs $\cE:=\{E_x\}_{x\in\cX}$ on $\cQ$.

We now state the non-adaptive quantum version of
Proposition~\ref{lem:guessingprobability}. It is a direct consequence of the reasoning above.
\newtheorem*{varlemmaguessing}{Proposition~\ref{lem:guessingprobability}$^\prime$}
\begin{varlemmaguessing}
Let $\bigExt:\cX\times\cY\rightarrow\cZ$ be a strong
$(k,\varepsilon)$-extractor, and let $\cF$ be a {\rm POVM} on
$\cQ$. Then for all cq-states $\rho_{XQ}$ with
\[
\hguessn{X}{Q}\geq k+\log\textfrac{1}{\varepsilon}\ ,
\]
we have
\[
\nonun{\rho}{\bigExt(X,Y)}{Y\cF(Q)}\leq 2\varepsilon\ .
\]
\end{varlemmaguessing}

The following is the quantum analogue of
Lemma~\ref{lem:wolflemmaclassical} (its proof can again be found in
Appendix \ref{app:a}). It states that a short
additional piece of classical information $W$ does not help much in
guessing $X$ if the quantum system $Q$ depends only on $X$. Again, additional independent information $V$ does not help either.
\newtheorem*{varwolflemma}{Lemma~\ref{lem:wolflemmaclassical}$^\prime$}
\begin{varwolflemma}
Consider a cccq-state $\rho_{XVWQ}$ with
$\rho_{XV}=\rho_X\otimes\rho_V$ and $VW\leftrightarrow
X\leftrightarrow Q$. Then
\[
\hguessn{X}{VWQ}\geq \hguessn{X}{Q}-H_0(W)\
\]
and with probability at least $1-\varepsilon$ over $(v,w)\leftarrow
P_{VW}$, we have
\[
\hguesscond{X}{Q}{V=v,W=w}\geq
\hguessn{X}{Q}-H_0(W)-\log\textfrac{1}{\varepsilon}\ .
\]
\end{varwolflemma}

We now state  more precisely what we are aiming to prove about strong extractors. Note that Lemma~\ref{lem:guessingprobability}$^\prime$ only gives a weak security guarantee for the extracted bits $\bigExt(X,Y)$ -- they are only shown to be secure against an adversary who measures his quantum state {\em before} receiving $Y$. To discuss the stronger type of security we aim for, we first state the  definition of the non-uniformity in the quantum case.
\begin{definition}\label{def:quantumextractor}
Let $\rho_{ZQ}$ be an arbitrary cq-state on $\cZ\otimes\cQ$. The
{\em non-uniformity $\nonun{\rho}{Z}{Q}$ of $Z$ given $Q$} is
defined as
\[
\nonun{\rho}{Z}{Q}:=\|\rho_{ZQ}-\rho_{\uniform_\cZ}\otimes \rho_Q\|\ ,
\]
where $\rho_{\uniform_\cZ}$ denotes the completely mixed state on $\cZ$.
\end{definition}
We describe a few basic properties of this definition in Appendix~\ref{app:properties}. In a cryptographic setting, the condition $\nonun{\rho}{Z}{Q}\leq \varepsilon$ for some small $\varepsilon$ means that the key $Z$ is secure in a setting where $Q$ is controlled by the adversary; as explained in~\cite{RenKoe05} (see also \cite{Renner05,KRBM05}), such a key is, with probability at least $1-\varepsilon$, equivalent to a perfectly secure key.

In the sequel, we aim to show that $\nonun{\rho}{\bigExt(X,Y)}{YQ}$ is small for certain strong extractors $\bigExt$ and appropriate parameters. This means that the extracted bits are secure even if the adversary is given $Y$ in addition to his quantum system.

In the next section, we will show that for extractors with binary output, 
the quantity of interest can  in fact be bounded by considering an adversary whose strategy does not depend on $Y$, i.e., he performs a measurement independent  of $Y$ as in Lemma~\ref{lem:guessingprobability}$^\prime$. We then use this result in Section \ref{sec:nonbinary} to construct strong extractors that
output several bits.

\section{Extractors with binary output\label{sec:binary}}

We will first sketch the arguments in this section. For an extractor $\Ext:\cX\times\cY\rightarrow \sbin$ with binary output the non-uniformity of the extracted bit $Z=\Ext(X,Y)$ given $Y$ and the quantum system $Q$ can be directly related to the success probability in distinguishing two quantum states $\rho_0^y$ and $\rho_1^y$, for each $y\in\cY$. For a given $y$ these are the (generally mixed) states of the adversary, conditioned on the  extracted bit being $0$ or $1$, respectively. We modify an argument by Barnum and Knill~\cite{BK:reversing}  to bound the optimal success probability in  distinguishing $\rho_0^y$ from $\rho_1^y$ for a given $y$ in terms of
the success probability  resulting from the use of a pretty good measurement~\cite{hausladenwootters94} $\cE^y_{pgm}$. On the other hand, we will show that there exists a POVM $\cF$ which refines all the pretty good measurements $\{\cE^y_{pgm}\}_{y\in\cY}$ simultaneously; i.e., the outcome of the measurement $\cE^y_{pgm}$ can be obtained by applying $\cF$ and classical post-processing. This POVM $\cF$ is a pretty good measurement  defined by the states $\{\rho_x\}_{x\in\cX}$, the  conditional states of $Q$ given $X=x$, or, in the bounded storage model, the states that the adversary prepares upon seeing $X$. Since the refined measurement ${\cal F}$ does not depend on $y$, we know that it cannot be superior to any classical strategy, see Proposition~\ref{lem:guessingprobability}$^\prime$, and we obtain the main
result of this section, Theorem \ref{thm:mainquantumbinary}.

In the next lemma we bound the non-uniformity $\nonun{\rho}{Z}{Q}$
of a cq-state $\rho_{ZQ}:=\sum_{z\in\sbin} p_z\proj{z}\otimes\rho_z$
with binary classical part using a pretty good measurement.

\begin{lemma}\label{lem:prettygoodmeasurementboundnon}
Let $\rho_{ZQ}:=\sum_{z\in\sbin} p_z\proj{z}\otimes\rho_z$ be a
cq-state with binary classical part. Then
\begin{align}
\nonun{\rho}{Z}{Q}\leq \sqrt{2\nonun{\rho}{Z}{\cE_{pgm}(Q)}}+d(Z)\ ,
\label{eq:pgmbound}
\end{align}
where $\cE_{pgm}$ is the pretty good measurement defined by
$\rho_{ZQ}$, i.e. the {\rm POVM} elements of this measurement are
$E_z:=p_z \rho_Q^{-\textfrac{1}{2}} \rho_z
\rho_Q^{-\textfrac{1}{2}}$ for $z\in\sbin$.
\end{lemma}
\begin{proof}
By definition
\begin{align}\label{eq:nonunbino}
\nonun{\rho}{Z}{Q}=\sum_{z=0}^1\|p_z\rho_z-\frac{1}{2}\rho_Q\|=\|p_0
\rho_0-p_1 \rho_1\|,
\end{align}
Let $\Delta:=p_0\rho_0-p_1\rho_1$ and let $\Delta=:A^+-A^-$ with
$A^+\geq 0$, $A^-\geq 0$ be the decomposition of $\Delta$ into a
nonnegative and a negative part. Then
\begin{align*}
\|p_0\rho_0-p_1\rho_1\|&=\frac{1}{2}(\tr(A^+)+\tr(A^-))\\
&=\tr(A^+)-\frac{1}{2}\tr(\Delta) \\
&=\tr(\mathbb{P}\Delta)-\frac{1}{2}(p_0-p_1),
\end{align*}
where $\mathbb{P}$ is the projector onto the support of $A^+$. We
will do some work to show that
\begin{align}\label{eq:firsteqneeded}
\tr(\mathbb{P}\Delta)\leq \sqrt{2\nonun{\rho}{Z}{\cE_{pgm}(Q)}}\ ,
\end{align}
where $\cE_{pgm}$ is the pretty good measurement that
distinguishes $\rho_1$ and $\rho_1$. By noting that
\begin{align*}
-\frac{1}{2}(p_0-p_1) \leq \frac{1}{2}|p_0-p_1|&\leq \frac{1}{2}\left(|p_0-\frac{1}{2}|+|p_1-\frac{1}{2}|\right)\nonumber\\
&=d(Z)\ ,
\end{align*}
we obtain the desired result, Eq.~\eqref{eq:pgmbound}. Consider thus the quantity
$\tr(\mathbb{P}\Delta)$. We can bound
\begin{align}
\tr(\mathbb{P}\Delta)\leq \sqrt{\tr(A^\dagger A)\tr(B^\dagger B)}\
\label{eq:cauchysc}
\end{align}
by applying the operator Cauchy-Schwarz inequality to the operators
\begin{align*}
A&:=\rho_Q^{\textfrac{1}{4}}\mathbb{P}\rho_Q^{\textfrac{1}{4}}\\
B&:=\rho_Q^{-\textfrac{1}{4}}\Delta\rho_Q^{-\textfrac{1}{4}}.
\end{align*}
But
\begin{align}
\tr(A^\dagger A) &=\tr(\rho_Q^{\textfrac{1}{2}}\mathbb{P}\rho_Q^{\textfrac{1}{2}}\mathbb{P})\nonumber\\
&\leq \tr(\rho_Q^{\textfrac{1}{2}}\mathbb{P}\rho_Q^{\textfrac{1}{2}})\nonumber\\
&\leq \tr(\rho_Q)=1\ \label{eq:adagad}
\end{align}
where we used the fact that $\mathbb{P}\leq \id$ and the fact that
$\rho_Q^{\textfrac{1}{2}}\mathbb{P}\rho_Q^{\textfrac{1}{2}}$ is
nonnegative. On the other hand, by the definition of the pretty good
measurement $\cE_{pgm}=\{E_0,E_1\}$ we have
\begin{align}
\tr(B^\dagger B)&=\tr(\rho_Q^{-\textfrac{1}{2}}\Delta\rho_Q^{-\textfrac{1}{2}}\Delta)\nonumber\\
&=\tr(E_0\Delta)-\tr(E_1\Delta)\nonumber\\
&=P_{succ}(\cE_{pgm})-p_1\tr(E_0\rho_1)-p_0\tr(E_1\rho_0)\nonumber\\
&=2 P_{succ}(\cE_{pgm})-1\ .\label{eq:psuccdefre}
\end{align}
Here we have used 
the definition of the success probability $P_{succ}(\{E_0,E_1\}):=p_0\tr(E_0\rho_0)+p_1\tr(E_1\rho_1)$, and the fact that $E_0+E_1=\id$ and $p_0+p_1=1$ in the
last step. Note that probability of success
$P_{succ}(\cE)$ for a fixed POVM $\cE$ is the same as the
probability of successfully distinguishing an instance drawn from
the distribution $\cE(\rho_0)$ and $\cE(\rho_1)$, respectively, with
a priori probabilities $p_0$ and $p_1$. Now we invoke Helstrom's
theorem~\cite{Helstrom76} which says that the success probability of
distinguishing two quantum states $\sigma_0$ and $\sigma_1$ with
priors $p_0$ and $p_1$ using an optimal POVM $\cE_{opt}$  is equal to $P_{succ}(\cE_{opt})=\frac{1}{2}+\|p_0\sigma_0-p_1\sigma_1\|$. We apply this theorem for $\sigma_z=\cE_{pgm}(\rho_z)$ and write $P_{succ}(\cE_{pgm})=\frac{1}{2}+\nonun{\rho}{Z}{\cE_{pgm}(Q)}$~(cf. Eq.~\eqref{eq:nonunbino}). Combining 
this with Eqs.~\eqref{eq:cauchysc},~\eqref{eq:adagad} and~\eqref{eq:psuccdefre} yields Eq.~\eqref{eq:firsteqneeded}, as desired.
\end{proof}
Now the goal is to bound the non-uniformity
$\nonun{\rho}{\Ext(X,Y)}{YQ}$ for extractors $\Ext$ with binary output when
$Q$ is a quantum system which depends on $X$. For this we consider
the cccq-state $\rho_{ZXYQ}\equiv\rho_{\Ext(X,Y)XYQ}$ which has the form
\begin{align}\label{eq:rhoxxdef}
\rho_{ZXYQ}=\sum_{y,x,z=\Ext(x,y)}P_{X}(x)P_Y(y)\proj{xyz}\otimes\rho_x\
,
\end{align}
where $P_Y(y)=\frac{1}{|\cY|}$ for every $y\in\cY$.
For this state one can express the non-uniformity $\nonun{\rho}{Z}{YQ}$ as (cf. Eq.~\eqref{eq:rhosigmadistance})
\begin{align}\label{eq:writeout}
\ExpE{y\leftarrow P_Y} \bigl[\sum_{z
\in\{0,1\}} \|\sum_x P_{Z|X=x,Y=y}(z) P_X(x)\rho_x-\frac{1}{2}\rho_Q\|\bigr]\ .
\end{align}
Note that $P_{Z|X=x,Y=y}(z)$ is $1$ or $0$, depending on whether or not $\Ext(x,y)=z$. It is straightforward to verify that
\begin{align}
p^y_z\rho^y_z &:=\sum_{x\in\cX} P_{Z|Y=y}(z)P_{X|Y=y,Z=z}(x)\rho_x\nonumber\\
&=\sum_{x\in\cX} P_{Z|X=x,Y=y}(z)P_{X}(x)\rho_x\ ,\label{eq:POVMusefl}
\end{align}
where we introduced for each $y\in\cY$ and $z\in\sbin$ the density matrix
\begin{align}
\rho^y_z:=\sum_{x\in\cX} P_{X|Y=y,Z=z}(x)\rho_x\ ,  \label{eq:defrhos}
\end{align}
with the normalising factor
\begin{align}\label{eq:defps}
p^y_z:=P_{Z|Y=y}(z)=\Pr_{x\leftarrow X}[ \Ext(x,y)=z]\ .
\end{align}
The state $\rho^y_z$ is the state of $Q$ conditioned on $\Ext(X,y)=z$; for any given $y\in\cY$, the two states $\rho^y_0$ and $\rho^y_1$ have a priori probabilities  $p^y_z$ with $z\in\sbin$. From this definition, it is clear that 
\begin{align}
\sum_z p^y_z \rho^y_z=\rho_Q\ , \label{eq:indepy}
\end{align}
which is {\em independent} of $y$.  This observation will be
essential in the proof of the following theorem.

Applying Helstrom's theorem gives an intuitive interpretation of the quantity of interest (which we state, but do not need later in the proof): $\frac{1}{2}+\nonun{\rho}{\Ext(X,Y)}{YQ}$ is the maximal average success probability when distinguishing $\rho^y_0$ and $\rho^y_1$ with a priori probabilities $p^y_0$ and $p^y_1$, over random $y\leftarrow P_Y$. This follows by combining Eq.~\eqref{eq:indepy} with Eq.~\eqref{eq:writeout} and Eqs.~\eqref{eq:defrhos},~\eqref{eq:defps}.

 We are ready to
derive the main result of this section:
\begin{theorem}\label{thm:mainquantumbinary}
Let $\Ext:\cX\times\cY\rightarrow\sbin$ be a strong
$(k,\varepsilon)$-extractor. Then for all $\rho_{XQ}$ with
\[
\hguessn{X}{Q}\geq k+\log\textfrac{1}{\varepsilon}\ ,
\]
we have
\[
\nonun{\rho}{\Ext(X,Y)}{YQ}\leq 3\sqrt{\varepsilon}\ ,
\]
where $\rho_{YXQ}:=\rho_{\uniform_{\cY}}\otimes\rho_{XQ}$.
\end{theorem}

\begin{proof}
By Eq.~\eqref{eq:rhosigmadistance} (cf. Eq.~\eqref{eq:expynonuniformity}) we can express
\begin{align}
\nonun{\rho}{\Ext(X,Y)}{YQ}=\ExpE{y\leftarrow
P_Y}[\nonun{\rho}{\Ext(X,y)}{Q}]\ .\label{eq:expresscond}
\end{align}
We can apply the pretty good measurement bound of Lemma
\ref{lem:prettygoodmeasurementboundnon} for each $y\in\cY$ 
to the state $\rho_{\Ext(X,y)Q}=\sum_{z\in\sbin} p^y_z \proj{z}\otimes\rho^y_z$,
where the
density matrices $\rho_z^y$ and their associated probabilities
$p_z^y$ are defined in Eqs.~\eqref{eq:defrhos} and \eqref{eq:defps}.
We get
\[
\nonun{\rho}{\Ext(X,y)}{Q}\leq
\sqrt{2\nonun{\rho}{\Ext(X,y)}{\cE^y_{pgm}(Q)}}+d(\Ext(X,y))
\]
for every $y\in\cY$. Taking the expectation over $y\leftarrow P_Y$
again and using the convexity of the square root gives
\begin{align}
\nonun{\rho}{\Ext(X,Y)}{YQ} &\leq
\sqrt{\ExpE{y\leftarrow P_Y}[2\nonun{\rho}{\Ext(X,y)}{\cE^y_{pgm}(Q)}]}\nonumber\\
&\qquad \qquad+\nonun{\rho}{\Ext(X,Y)}{Y} \label{eq:midway}
\end{align}
by Eq.~\eqref{eq:rhosigmadistance} (see also Eq.~\eqref{eq:nonuniformclassicalaverage}). Since
\[
H_\infty(X)\geq \hguessn{X}{Q}\ ,
\]
the second term in Eq.~\eqref{eq:midway} is upper bounded by
$\varepsilon$. Let us now consider the details of the pretty good
measurement $\cE^y_{pgm}$. The measurement $\cE^y_{pgm}=\{E^y_z\}_{z\in\sbin}$
is determined by the POVM elements
\begin{align}\label{eq:eyzdef}
E^y_z:=p^y_z(G^y)^{-\textfrac{1}{2}}\rho^y_z(G^y)^{-\textfrac{1}{2}}
\end{align}
where, as argued above (Eq.~\eqref{eq:indepy}), 
\begin{align}\label{eq:gydf}
G^y=\sum_{z\in\cZ} p^y_z\rho^y_z=\rho_Q\ .
\end{align}
is independent of $y$. This fact allows us to define a new pretty good
measurement $\cF$ which does not depend on $y$, but is equally good
or better in estimating $Z$ from $Q$ and $Y$. This new pretty good
measurement $\cF=\{F_x\}_{x\in\cX}$ has POVM elements
\[
F_x:=P_X(x)\rho_Q^{-\textfrac{1}{2}}\rho_x\rho_Q^{-\textfrac{1}{2}}\ .
\]
Expressed differently, $\cF$ is simply the pretty good measurement defined by the ensemble $\{P_X(x),\rho_x\}$. From Eqs.~\eqref{eq:POVMusefl},~\eqref{eq:eyzdef}  and~\eqref{eq:gydf} above one can see that
\begin{align}
E^y_z&=\sum_{x\in\cX} P_{Z|X=x,Y=y}(z)F_x\label{eq:eyzeronedef}
\end{align}
In other words, the results of the measurements $\{\cE^y_{pgm}\}_{y\in\cY}$ can in
fact be obtained by first estimating $x$ by measuring the quantum system
$Q$ with $\cF=\{F_x\}_{x\in \cX}$. Then we infer $z$ for a given $y$
by computing $z=\Ext(x,y)$. On a more technical level, one
needs to show that  for every $y\in\cY$ the non-uniformity given the measurement outcome
of the  measurement $\cE^y_{pgm}$ is smaller than or equal to the non-uniformity given the outcome of the refined
measurement ${\cal F}$. We have summarized these technical details
in Lemma~\ref{lem:refinementlemma} proved in Appendix~\ref{app:refine}. Formally, we have
\[
\nonun{\rho}{\Ext(X,y)}{\cE^y_{pgm}(Q)}\leq \nonun{\rho}{\Ext(X,y)}{\cF(Q)}\ .
\]
Taking the expectation over $y\leftarrow P_Y$ gives (cf. Eq.~\eqref{eq:expynonuniformity})
\[
\ExpE{y\leftarrow P_Y}[\nonun{\rho}{\Ext(X,y)}{\cE^y_{pgm}(Q)}] \leq
\nonun{\rho}{\Ext(X,Y)}{Y\cF(Q)}\ .
\] Since ${\cal F}$ does not depend
on $y$ we have reduced our problem to the simple scenario where the
quantum system is measured before the adversary obtains $y$.
Thus we can apply Proposition~\ref{lem:guessingprobability}$^\prime$,
\begin{align}\label{eq:alphabound}
\nonun{\rho}{\Ext(X,Y)}{Y\cF(Q)}\leq 2\varepsilon\ .
\end{align}
We conclude with Eq.~\eqref{eq:midway} that
\[
\nonun{\rho}{\Ext(X,Y)}{YQ}\leq 2\sqrt{\varepsilon}+\varepsilon\ ,
\]
hence the claim follows.
\end{proof}

We now show that even if the adversary is given additional
information $V$ which is independent of $X$ and a short bit string
$W$ which might depend on $X$, the extracted bit looks secure. This
statement will be used below to prove that certain extractors which output several bits can also safely be used in a cryptographic context (cf. Theorem~\ref{thm:maintheorem}).

\begin{corollary}\label{cor:wolfextlemma}
Let $\Ext:\cX\times\cY\rightarrow \sbin$ be a strong $(k,\varepsilon)$-extractor. Let $\rho_{XVWQ}$ be a cccq-state with $\rho_{XV}=\rho_X\otimes\rho_V$, $VW\leftrightarrow X\leftrightarrow Q$ and
\[
\hguessn{X}{Q}\geq k+H_0(W)+2\log\textfrac{1}{\varepsilon}\ .
\]
Then
\[
\nonun{\rho}{\Ext(X,Y)}{YVWQ}\leq 4\sqrt{\varepsilon}\ ,
\]
where $\rho_{YXVWQ}:=\rho_{\uniform_{\cY}}\otimes\rho_{XVWQ}$.
\end{corollary}

\begin{proof}
Let $\alpha:=\nonun{\rho}{\Ext(X,Y)}{YVWQ}$ be the quantity of interest. Then by Eq.~\eqref{eq:rhosigmadistance} (see also Eq.~\eqref{eq:nonuniformclassicalaverage}),
\[
\alpha=\ExpE{(v,w)\leftarrow P_{VW}}\bigl[\nonun{\rho}{\Ext(X,Y)}{YQ|V=v,W=w}\bigr]\ ,
\]
where the term in brackets is the non-uniformity of $\Ext(X,Y)$ with respect to the conditional state $\rho_{XQ|V=v,W=w}$.
By  Lemma~\ref{lem:wolflemmaclassical}$^\prime$, we have
\begin{align}\label{eq:hguesscbd}
\hguesscond{X}{Q}{V=v,W=w}\geq k+\log\textfrac{1}{\varepsilon}\
\end{align}
with probability at least $1-\varepsilon$ over random $(v,w)\leftarrow P_{VW}$.
For any $(v,w)$ for which Eq.~\eqref{eq:hguesscbd} is satisfied, we have
\[
\nonuncond{\rho}{\Ext(X,Y)}{YQ}{V=v,W=w}\leq 3\sqrt{\varepsilon}\
\]
by Theorem~\ref{thm:mainquantumbinary}.  Thus
\[
\alpha\leq 3\sqrt{\varepsilon}+\varepsilon\ ,
\]
and the claim follows.
\end{proof}

\section{Extractors with nonbinary output\label{sec:nonbinary}}

In this section we will consider strong extractors which output several bits. We first show how to use
independent seeds $y_1, \ldots, y_m$ to extract $m$ bits. The
security of the extracted bits in the quantum setting will follow from
applying our bound for binary extractors, Theorem~\ref{thm:mainquantumbinary} in combination with a quantum version of
the so-called hybrid argument. By a similar technique, we will show how to extract more bits under stronger assumptions. Let us first discuss the hybrid argument.

Consider a cq-state of the form $\rho_{ZQ}$, where $Z=(Z_1,\ldots,Z_m)$ is an $m$-bit string. We aim to find a bound on
$\nonun{\rho}{Z}{Q}$ in terms of non-uniformities of binary random variables.

By definition, we have
\begin{align*}
\nonun{\rho}{Z}{Q}&=\|\rho_{ZQ}-\rho_{\uniform_{\sbin}}^{\otimes
m}\otimes \rho_Q\|\ .
\end{align*}
Let us define for $i=0,\ldots,m$ the states
\[
\rho^{(i)}:=\rho_{\uniform_{\sbin}}^{\otimes m-i}\otimes
\rho_{Z^{i}Q}\
\]
on $\sbin^m\otimes\cQ$, where we use the abbreviation
$z^i:=(z_1,\ldots,z_i)$ to refer to the first $i$ bits of
$z\in\sbin^m$. Clearly, we have $\rho^{(m)}=\rho_{ZQ}$ and
$\rho^{(0)}=\rho_{\uniform_{\sbin}}^{\otimes m}\otimes \rho_Q$. We
use the ``telescoping'' sum
\[
\rho^{(0)}-\rho^{(m)}=\sum_{i=0} ^{m-1} \rho^{(i)}-\rho^{(i+1)}\ ,
\]
which by the triangle inequality implies that
\[
\nonun{\rho}{Z}{Q}\leq \sum_{i=0} ^{m-1}\|\rho^{(i+1)}-\rho^{(i)}\|\
.
\]
But
\begin{align*}
\|\rho^{(i+1)}-\rho^{(i)}\|&=\|\rho_{\uniform_{\sbin}}^{\otimes m-i-1}\otimes \rho_{Z^{i+1}Q}-\rho_{\uniform_{\sbin}}^{\otimes m-i}\otimes \rho_{Z^{i}Q}\|\\
& =\|\rho_{Z^{i+1}Q}-\rho_{\uniform_{\sbin}}\otimes\rho_{Z^{i}Q}\|
\\
&=\|\rho_{Z_{i+1}Z^iQ}-\rho_{\uniform_{\sbin}}\otimes\rho_{Z^iQ}\|\
\end{align*}

We thus arrive at the following conclusion
\begin{align}
\nonun{\rho}{Z}{Q}\leq \sum_{i=0} ^{m-1}
\nonun{\rho}{Z_{i+1}}{Z^iQ}\ .~\label{eq:hybrid}
\end{align}

Let us now state and prove the main theorem.
\begin{theorem}\label{thm:maintheorem}
Let $\Ext:\cX\times\cY\rightarrow \sbin$ be a strong
$(k,\varepsilon)$-extractor, and let
\begin{align*}
\bigExt^m:\cX\times\cY^{m}&\rightarrow \sbin^m\\
(x,y_1,\ldots,y_m)&\mapsto (\Ext(x,y_1),\ldots,\Ext(x,y_m))\ .
\end{align*}
Then for all cq-states $\rho_{XQ}$ with
\begin{align}\label{eq:boundedstorhguesscond}
\hguessn{X}{Q}\geq k+m+2\log\textfrac{1}{\varepsilon}\ ,
\end{align}
we have \
\[
\nonun{\rho}{\bigExt^m(X,Y^m)}{Y^mQ}\leq 4m\sqrt{\varepsilon}\ ,
\]
where $\rho_{Y^mXQ}:=\rho_{\uniform_{\cY^m}}\otimes\rho_{XQ}$.
\end{theorem}
\begin{proof}
We use Eq.~(\ref{eq:hybrid}) to get
\begin{align}\label{eq:triangleinapplyied}
\nonun{\rho}{\bigExt^m(X,Y^m)}{Y^mQ}\leq \sum_{i=0} ^{m-1}
\nonun{\rho}{Z_{i+1}}{Z^iY^mQ}\ ,
\end{align}
where $Z^m\equiv \bigExt^m(X,Y^m)$. Observe that
$(Y_{i+2},\ldots,Y_m)$ is independent of $Z^{i+1}Y^{i+1}Q$, which
by Eq.~\eqref{eq:nonuniformindependent}
gives
\begin{align}
\nonun{\rho}{Z_{i+1}}{Z^iY^mQ}=\nonun{\rho}{Z_{i+1}}{Z^iY^{i+1}Q}\
.\label{eq:triangleinapplyiedsecond}
\end{align}
But
\begin{align}
\nonun{\rho}{Z_{i+1}}{Z^iY^{i+1}Q}&=\nonun{\rho}{Z_{i+1}}{Y_{i+1}Z^iY^iQ}\nonumber\\
&=\nonun{\rho}{\Ext(X,\tilde{Y})}{\tilde{Y}\bigExt^i(X,Y^i)Y^iQ}\nonumber
\end{align}
where $\tilde{Y}\equiv Y_{i+1}$. Applying Corollary~\ref{cor:wolfextlemma}
to $(V,W)=(Y^i,\bigExt^i(X,Y^i))$ yields
\begin{align}\label{eq:triangleinapplyiedthird}
\nonun{\rho}{Z_{i+1}}{Z^iY^{i+1}Q}\leq 4\sqrt{\varepsilon}\
\end{align}
for every $i=0,\ldots,m-1$. 
We have made use of the fact that $H_0(W)=H_0(\bigExt^i(X,Y^i)) \leq m$ by definition. The claim then follows
from Eqs.~\eqref{eq:triangleinapplyied},~\eqref{eq:triangleinapplyiedsecond}
and~\eqref{eq:triangleinapplyiedthird}.
\end{proof}
In the next section, we study the implications of Theorem~\ref{thm:maintheorem} for the bounded storage model.  We will see that the bound on the storage of the adversary translates into an upper bound on the guessing probability, as required (cf. Eq.~\eqref{eq:boundedstorhguesscond}). We will then give a concrete example of an extractor for the bounded storage model with quantum adversaries.

Before continuing, however, let us point out that in certain situations, we can use the hybrid argument to show that the seed $Y$ can be reused several times. This gives more efficient randomness extractors (under stronger assumptions about the inital cq-state $\rho_{XQ}$). Following similar terminology in the literature on extractors, we introduce the following notion.
\begin{definition}
A cq-state $\rho_{XQ}$ where $X=(X_1,\ldots,X_m)$ consists of $m$ parts is a {\em $k$-blockwise state} if for all $i=0,\ldots,m-1$
\[
\hguessn{X_{i+1}}{X^iQ}\geq k\ .
\]
\end{definition}
We will now show how to extract multiple bits from such a cq-state by reusing the seed. This is interesting for several reasons. First, $k$-blockwise states arise naturally in realistic situations such as the bounded storage model. We will discuss this in more detail below (cf. Section~\ref{sec:independentrandomizers}). Second, extractors for $k$-blockwise probability distributions are often used to construct (classical) extractors by transforming the input distribution to a $k$-blockwise distribution. It might therefore be possible to obtain extractor constructions for the quantum case using similar lines of reasoning.
\begin{theorem}\label{thm:maintheoremblockwise}
Let $\Ext:\cX\times\cY\rightarrow \sbin$ be a strong
$(k,\varepsilon)$-extractor, and let
\begin{align*}
\tilde{\bigExt}^L:\cX^L\times\cY^m&\rightarrow \sbin^{Lm}\\
(x_1,\ldots,x_L,y)&\mapsto (\bigExt^m(x_1,y),\ldots,\bigExt^m(x_L,y))\ ,
\end{align*}
where $\bigExt^m:\cX\times\cY^m\rightarrow \sbin^m$ is defined as in Theorem~\ref{thm:maintheorem}.
Then 
\[
\nonun{\rho}{\tilde{\bigExt}^L(X^L,Y)}{YQ}\leq 4Lm\sqrt{\varepsilon}\ ,
\]
for all $(k+m+2\log\textfrac{1}{\varepsilon})$-blockwise states $\rho_{XQ}$ on $\cX^L\otimes\cQ$,
where $\rho_{YXQ}:=\rho_{\uniform_{\cY}}\otimes\rho_{XQ}$.
\end{theorem}
\begin{proof}
With Eq.~(\ref{eq:hybrid}) we get
\begin{align}
\nonun{\rho}{\tilde{\bigExt}^L(X^L,Y)}{YQ}\leq \sum_{i=0} ^{L-1}
\nonun{\rho}{Z_{i+1}}{Z^iYQ}\ ,
\end{align}
where $Z^L\equiv \tilde{\bigExt}^L(X^L,Y)$. Since $(Z^i,Y)$ is a
function of $(X^i,Y)$ and since applying functions does not increase
the trace distance, we
obtain
\begin{align}
\nonun{\rho}{\tilde{\bigExt}^L(X^L,Y)}{YQ}\leq \sum_{i=0} ^{L-1}
\nonun{\rho}{Z_{i+1}}{X^iYQ}\ .
\end{align}
But $\nonun{\rho}{Z_{i+1}}{X^iYQ}=\nonun{\rho}{\bigExt^m(X_{i+1},Y)}{YX^iQ}$, and
$\rho_{YX_{i+1}X^iQ}=\rho_{\uniform_{\cY}}\otimes\rho_{X_{i+1}X^iQ}$. Moreover,
\[
\hguessn{X_{i+1}}{X^iQ}\geq k+m+2\log\textfrac{1}{\varepsilon}\ .
\]
by assumption. Thus we can apply Theorem~\ref{thm:maintheorem} and the claim follows.
\end{proof}

\section{The Bounded Storage Model with a Quantum Adversary\label{sec:boundedstorage}}

\subsection{Bounded storage, guessing entropy and extractors}
In the classical version of the bounded storage model, the security of the extracted bits is a direct consequence of the property of the extractor given in  Proposition~\ref{lem:guessingprobability} and the fact that an adversary has limited information about $X$. The latter fact is expressed by the following well-known proposition, whose proof we omit, as it is trivial. It states that an adversary who has $H_0(E)$ bits of storage can not predict $X$ well.

\begin{proposition}\label{lem:wolflemma}
Let $P_{XE}$ be an arbitrary distribution. Then
\begin{align*}
\hguessn{X}{E}\geq H_\infty(X)-H_0(E)\ .
\end{align*}
\end{proposition}
Together with Lemma~\ref{lem:guessingprobability}, it follows that a strong $(k, \epsilon)$-extractor has the property that $\nonun{\rho}{\Ext(X,Y)}{YE}\leq 2\epsilon$ for all $P_{XE}$ with 
\[
H_\infty(X) \geq k+H_0(E)+\log\textfrac{1}{\varepsilon}\ .
\]
Thus the security of the extracted key  can be directly derived from the strong extractor property and the bounded-storage assumption. The main challenge is to construct strong extractors which satisfy all the additional requirements for applicability in the bounded storage model (see Section~\ref{sec:concreteexample}).

What about quantum storage? We show that a similar reasoning applies; given an extractor which is characterized by the guessing-entropy $\hguessn{X}{Q}$, the storage bound  can be translated into a security guarantee. We first show that $X$ can not be guessed by measuring $Q$ when the number of qubits constituting $Q$ is limited.
\newtheorem*{varwolflemmasec}{Proposition~\ref{lem:wolflemma}$^\prime$}
\begin{varwolflemmasec}
Let $\rho_{XQ}$ be a cq-state. Then
\[
\hguessn{X}{Q}\geq H_\infty(X)-H_0(Q)\ .
\]
\end{varwolflemmasec}

\begin{proof}
Consider a {\rm POVM} $\cE:=\{E_x\}_{x\in\cX}$ on $\cQ$ that
maximizes the expression defining $\hguessn{X}{Q}$ (cf.
Eq~(\ref{eq:hguessQdef})).
 Then
\begin{align*}
2^{-\hguessn{X}{Q}}&= \sum_{x} P_X(x)\tr(E_x\rho_x)\\
&=\tr\Bigl(\bigl(\sum_x \proj{x}\otimes E_x\bigr)\rho_{XQ}\Bigr)\\
&\leq 2^{-H_\infty(XQ)}\tr\Bigl(\sum_x \proj{x}\otimes E_x\Bigr)\ .
\end{align*}
The statement then follows from the fact that
\[
\tr\Bigl(\sum_x \proj{x}\otimes E_x\Bigr)=\tr\Bigl(\sum_x E_x\Bigr)=\tr(\id_{\cQ})=2^{H_0(Q)}\ .
\]
\end{proof}

By combining Proposition~\ref{lem:wolflemma}$^\prime$ with
Theorem~\ref{thm:maintheorem}, we obtain a way of
constructing strong extractors for the bounded storage model in the
presence of quantum adversaries: the statement of Theorem~\ref{thm:maintheorem}  holds when Eq.~\eqref{eq:boundedstorhguesscond} is replaced by the weaker condition
\begin{align}\label{eq:hinfhzerocond}
H_\infty(X)\geq k+m+H_0(Q)+2\log\textfrac{1}{\varepsilon}\ .
\end{align}
Before applying  this result to obtain a concrete construction, 
let us elaborate on a recent example which shows that not every strong extractor yields secure bits in the quantum bounded storage model.
\begin{remark}\label{rem:gavkempdewolf}
Gavinsky, Kempe and de Wolf~\cite{GKW:sepcc} consider the function
\begin{align*}
\Ext:&\sbin^n\times \omega_n&\rightarrow &\qquad \sbin\\
&((x_1,\ldots,x_n),\{y_1,y_2\})&\mapsto&\qquad x_{y_1}\oplus x_{y_2}\ ,
\end{align*}
where $\oplus$ denotes bitwise addition modulo $2$ and 
where $\omega_n$ is the set of pairs $(y_1,y_2)$ of distinct indices $y_1,y_2\in\{1,\ldots,n\}$.  They then study the function $\bigExt^m$ restricted to the set $\sbin^n\times \Omega_m$, where $\Omega_m\subset \omega_n^m$ is the subset of disjoint $m$-tuples. Let us call this restriction $\tilde{\bigExt}_m$ and let $\tilde{Y}_m$ be uniform on~$\Omega_m$. In our terminology, they show the following. There is an
$\alpha\approx\textfrac{1}{\sqrt{\log n}}$ such that for large enough $n$ and
$m:=\alpha n$, the quantity $\nonun{P}{\tilde{\bigExt}_m(X,\tilde{Y}_m)}{\tilde{Y}_mE}$ is
small for any classical random variable $E$ with $H_0(E)\leq
\sqrt{n}$, whereas  $\nonun{P}{\tilde{\bigExt}_m(X,\tilde{Y}_m)}{\tilde{Y}_mQ}$ is large if $Q$ is quantum and $H_0(Q)$ is polylogarithmic in $n$.

This statement does not contradict Theorem~\ref{thm:maintheorem} which can not be applied in this situation. This is because the function $\tilde{E}_m$ does not have the required form. While Theorem~\ref{thm:mainquantumbinary} tells us that the difference between classical and quantum prior information is limited in the case of extractors with binary output, this example shows that the case of general extractors which output several bits is more subtle.
\end{remark}

\subsection{ Extractors for the bounded storage model:  an explicit example\label{sec:concreteexample}}
In this section, we give a concrete example of a function $\bigExt:\sbin^n\times\sbin^t\rightarrow \sbin^m$ which can be used in the bounded storage model in the presence of a quantum adversary. Let us first discuss what additional requirements such a function has to satisfy.

Typical parameters of the bounded storage model are as follows: For some $1\geq \alpha> \beta>0$,  $H_{\infty}(X)\geq \alpha n$ and $H_0(Q)\leq \beta n$. Here, the parameter $\alpha$ is called the {\em min-entropy rate}, whereas $\beta$ is referred to as the {\em storage rate}. The amount of memory available to the honest parties, Alice and Bob, on the other hand, is supposed to be much more limited. Typically, it is assumed that they have only $O(\log n)$ bits of storage. Expressed differently, the scheme should be secure even if the adversary is significantly more powerful than the participating honest parties.

The fact that Alice and Bob have only $O(\log n)$ bits of memory implies that the strong extractor must have seed length $\log |\cY|=t$ of that order. Moreover, the extractor has to be (efficiently) computable with limited memory. This is the case if $\bigExt$ is {\em $\ell$-local}, meaning that it only depends on a small number $\ell$ (instead of $n$)  physical bits of its first argument, where the $\ell$ bit locations are determined by the second argument. Note that a different solution to the latter problem was suggested by Lu~\cite{Lu02}, who considers so-called on-line computable functions.

Due to these requirements, finding explicit, efficiently computable constructions for the bounded-storage model is a rather intricate problem, which has been studied for some time~\cite{Maurer92b,ADR:bs,DziMau02,Lu02,Vadhan03}. 
Here we consider a construction by Vadhan.
By choosing the output to be a single bit,  Theorem~8.5 in~\cite{Vadhan03} gives an $\ell$-local strong $(k,\varepsilon)$-extractor $\Ext:\sbin^n\times\sbin^t\rightarrow\sbin$ with
\begin{align}
t &= \log n+O(\log\textfrac{1}{\varepsilon})\label{eq:vadhanfirst}\\
\ell &=\frac{3}{2}\frac{n}{k}+O(\log\textfrac{1}{\varepsilon})\label{eq:vadhansecond}
\end{align}
for every $\varepsilon>\exp (-\textfrac{n}{2^{O(\log^*n)}})$.

Suppose we want to achieve an error $\varepsilon$, using Theorem~\ref{thm:maintheorem}. Then the error for the one-bit-extractor $\Ext$ must be upper bounded by $(\textfrac{\varepsilon}{4m})^2$. Inserting this into Eqs.~\eqref{eq:vadhanfirst} and~\eqref{eq:vadhansecond} gives
\begin{corollary}\label{cor:vadhanquantum}
For any $\varepsilon>4m\exp (-\textfrac{n}{2^{O(\log^*n)}})$, there is an $\ell$-local function
\[
\bigExt:\sbin^n\times\sbin^t\rightarrow \sbin^m\ ,
\]
with
\begin{align*}
t&=m\log n+O(m\log m+m\log\textfrac{1}{\varepsilon})\\
\ell&=\frac{3}{2}\frac{nm}{k-m-2\log\textfrac{1}{\varepsilon}}+O(m\log m+m\log\textfrac{1}{\varepsilon})\ ,
\end{align*}
such that $\nonun{\rho}{\bigExt(X,Y)}{YQ}\leq \varepsilon$
for all  $\rho_{XQ}$ with
\[
H_{\infty}(X)-H_0(Q)\geq k\ ,
\]
where $\rho_{YXQ}=\rho_{\uniform_{\sbin^d}}\otimes \rho_{XQ}$.
\end{corollary}
In terms of the  min-entropy rate $\alpha$ and the storage-rate $\beta$, our  result implies that for any $\alpha> \beta$,
 there is an extractor which uses $O(m\log n+m\log\textfrac{1}{\varepsilon})$  bits of initial key, outputs $m$ bits with security $\varepsilon$, and reads $O(m\log m+m\log\textfrac{1}{\varepsilon})$ bits from the randomizer $X$. In comparison, the best known   classical construction uses $O(\log n+\log\textfrac{1}{\varepsilon})$ bits of key  and reads $O(m+\log\textfrac{1}{\varepsilon})$ from $X$.

\subsection{Independent randomizers\label{sec:independentrandomizers}}
In the so-called satellite scenario~\cite{Maurer92b}, the randomizer $X$ is assumed to consist of a sequence of random bits that are publicly broadcast in sequence. In this situation, it is clear that if we partition $X$ into blocks $X=(X_1,\ldots,X_L)$, the random variables corresponding to the blocks are independent. What is more interesting is that if the adversary is allowed to prepare a quantum system $Q$ adaptively, the resulting cq-state $\rho_{XQ}$ is a $k$-blockwise state. This is a consequence of the fact that taking the previous blocks $X^i$ into account when storing and retrieving information about $X_{i+1}$ does not help the adversary if $X_{i+1}$ is independent of $X^i$. We can express this formally by the following result, with the set $\cS$ corresponding to all states on a Hilbert space of limited dimension in the bounded storage model. 

\begin{lemma}
Let $P_{XX'}=P_X\cdot P_{X'}$ be a probability distribution of independent random variables and let $\cS$ be a set of states. Then
\begin{align}\label{eq:hgindeprandom}
\min_{\rho_{XX'Q}} \hguessn{X}{X'Q}\geq \min_{\rho_{XQ}}\hguessn{X}{Q}\ ,
\end{align}
where the minima are over all states of the form
\[
\rho_{XX'Q}=\sum_{x,x'}P_{XX'}(x,x')\proj{x}\otimes\proj{x'}\otimes\rho^{x'}_{x}\  
\]
with $\rho^{x'}_x\in\cS$ and 
\[
\rho_{XQ}=\sum_{x}P_{X}(x)\proj{x}\otimes\rho_{x}\ ,\qquad\ \rho_x\in\cS\ ,
\]
respectively.
\end{lemma}
\begin{proof}
Let $\{\rho^{x'}_x\}_{x,x'}$ be a family of states such that the corresponding state $\rho_{XX'Q}$ achieves the minimum on the l.h.s. of Eq.~\eqref{eq:hgindeprandom}. Then
\[
2^{-\hguessn{X}{X'Q}}=\ExpE{x'\leftarrow P_{X'}} [2^{-\hguesscond{X}{Q}{X'=x'}}]\ .
\]
But
\begin{align*}
2^{-\hguesscond{X}{Q}{X'=x'}}&=\max_{\{E_x\}_x} \sum_{x\in\cX}P_X(x)\tr(E_x\rho^{x'}_x)\\
&=2^{-\hguessn{X}{Q}}\ ,
\end{align*}
where the latter expression denotes the guessing entropy of $X$ given $Q$ in the state 
\[
\rho^{x'}_{XQ}=\sum_{x\in\cX}P_X(x)\proj{x}\otimes\rho^{x'}_x\ .
\]
The claim directly follows from this.
\end{proof}

If the randomizer $X$ consists of several independent parts
 $X=(X_1,\ldots,X_L)$ which satisfy $H_\infty(X_i)\geq H_0(Q)+k$ for all $i$, we can therefore use our hybrid construction (Theorem~\ref{thm:maintheoremblockwise}) in conjunction with Corollary~\ref{cor:vadhanquantum}.  As an example, consider the case where
 each of the blocks $X_i$ consists of  $n$ bits  with min-entropy rate $\alpha$. We then obtain an extractor $\bigExt:\sbin^{Ln}\times\sbin^t\rightarrow \sbin^{Lm}$ which uses $t=mO(\log n+\log L+\log\textfrac{1}{\varepsilon})$ bits of initial key, reads $mO(\log m+\log L+\log\textfrac{1}{\varepsilon})$ bits from $X$ and gives an $\varepsilon$-secure output in the presence of an adversary with storage rate $\beta <\alpha$. In particular, this construction can extend the key of the honest parties by more than the number of initial key bits. This implies that Alice and Bob end up with a longer key even if the adversary later learns the initial key $Y$.

\section{Tomography-based approach to general extractors\label{sec:tomography}}
The results of Section~\ref{sec:binary} imply that the
security of a single extracted bit is similar with respect to an adversary that has quantum instead of classical resources.

This is not true for general extractors which output several bits,
as shown in~\cite{GKW:sepcc} by an explicit counterexample. It is
however possible to give constructions that extract multiple bits
in a useful way, as we have shown in the previous section.

Which constructions give rise to ``useful'' extractors in
a quantum context? In this section, we elaborate on this 
question, showing that general extractors
can be used in the setting of privacy amplification,  {\em if} the adversary's memory is limited. Note that the setting of privacy amplification imposes
less stringent requirements on the extractor than the setting of the
bounded-storage model. Nevertheless, the only construction known to
work in the quantum setting has been two-universal
hashing~\cite{BenOr02,KoMaRe05,RenKoe05,Renner05}.

While two-universal hashing has the advantage that it extracts all the randomness present in the source (i.e., the number of extracted bits can be as large as $H_\infty(X)-2\log\textfrac{1}{\varepsilon}$), it requires a long seed $Y$. Viewed as an extractor, two-universal hashing has the form $\bigExt:\sbin^n\times\sbin^n\rightarrow \sbin^m$, i.e., the seed is of  the same length as the source $X$.
When applied to privacy amplification, this means that $n$ bits need to be communicated from Alice to Bob. We will show below that by considering general extractors the amount of communication can be reduced to approximately the number of qubits the adversary controls. This is important for applications such as the protocols proposed in~\cite{DFSS05}.

We use a measurement-based approach, which bounds the trace distance in
terms of the outcomes of a tomographic measurement in mutually
unbiased bases. More precisely, we will use the following lemma, whose proof is analogous to a proof in~\cite{divincenzo+:locking}.
\begin{lemma}\label{lem:mutuallyunbiasedsec}
Let $A$ be a hermitian operator on $\cQ$, where $d:=\dim\cQ=p^n$ is a prime power. Then there is {\rm POVM} $\cF$ such
that
\[
\|A\|\leq (d+1)\cdot \|\cF(A)\|\ .
\]
\end{lemma}

We can then show the following:
\begin{lemma}\label{lem:tomographyextr}
Let $\bigExt:\cX\times\cY\rightarrow \cZ$ be a strong
$(k,\varepsilon)$-extractor. Then
\begin{align}\label{eq:dexth0}
\nonun{\rho}{\bigExt(X,Y)}{YQ}\leq 4\cdot 2^{H_0(Q)}\cdot \varepsilon\
,
\end{align}
for all cq-states $\rho_{XQ}$ with
\[
\hguessn{X}{Q}\geq  k+\log\textfrac{1}{\varepsilon}\ ,
\]
and $2^{H_o(Q)}=p^m$ for some prime $p$.
\end{lemma}

\begin{remark} We point out that the condition on the dimension of
$Q$ can easily be dropped by using a different  measurement than the
one described in Lemma~\ref{lem:mutuallyunbiasedsec}, at the cost of
introducing an additional constant in the exponent on the r.h.s. of
Eq.~\eqref{eq:dexth0}.
\end{remark}

\begin{proof}
By definition and Eq.~\eqref{eq:rhosigmadistance},
\[
\nonun{\rho}{\bigExt(X,y)}{Q}=\sum_{z\in\cZ}
\|P_{\bigExt(X,y)}(z)\rho^z_y-\frac{1}{|\cZ|}\rho_Q\|\ ,
\]
where $\rho^z_y$ is the conditional state
$\rho^z_y:=\rho_{Q|\bigExt(X,y)=z}$
 for all $(y,z)\in\cY\times\cZ$. By Lemma~\ref{lem:mutuallyunbiasedsec}, we get
\begin{align*}
&\nonun{\rho}{\bigExt(X,y)}{Q}\\
&\qquad\leq 2^{H_o(Q)+1}\sum_{z\in\cZ}\|P_{\bigExt(X,y)}(z)\cF(\rho^z_y)-\frac{1}{|\cZ|}\cF(\rho_Q)\|\ ,
\end{align*}
and thus by taking the expectation over $y\leftarrow P_Y$ with Eq.~\eqref{eq:rhosigmadistance} (see also Eq.~\eqref{eq:expynonuniformity})
\[
\nonun{\rho}{\bigExt(X,Y)}{YQ}\leq
2^{H_o(Q)+1}\,\nonun{\rho}{\bigExt(X,Y)}{Y\cF(Q)}\ .
\]
The claim then follows from
Proposition~\ref{lem:guessingprobability}$^\prime$.
\end{proof}
This lemma shows that 
in principle, any strong $(k,\varepsilon)$-extractor with suitable parameters can be used for privacy amplification. We
illustrate this using a construction by Srinivasan and Zuckerman~\cite{srinizuck} for simplicity, but  we point out that using constructions from~\cite{reingold00extracting}, it is  possible to reduce the randomness required for privacy amplification even further. They give an efficiently computable strong $(k,\varepsilon)$-extractor $\bigExt:\sbin^n\times\sbin^t\rightarrow\sbin^m$  for any $k,m,\varepsilon$ with $k\geq m+2\log\textfrac{1}{\varepsilon}+2$, where $t=2(k+m)+O(\log n)$. Applying this to a situation where the adversary is given at most $d\geq H_0(Q)$ qubits of storage, we obtain an efficiently computable function $\bigExt:\sbin^n\times\sbin^t\rightarrow\sbin^m$ which uses only
\[
t=4(d+m+\log\textfrac{1}{\varepsilon}+3)+O(\log n)
\]
bits of seed and satisfies $\nonun{\rho}{\bigExt(X,Y)}{YQ}\leq \varepsilon$ whenever 
\[
H_\infty(X)\geq m+4d+3\log\textfrac{1}{\varepsilon}+8\ .
\]
For certain parameters, this construction is more efficient in terms of the seed length $t$ than the local extractor described in Corollary~\ref{cor:vadhanquantum}.

\section{Conclusions \label{sec:conclusions}}
While Holevo's celebrated theorem implies that $n$ quantum bits can not be used to store more than $n$ classical bits reliably, this result is in general not applicable in cryptography, where even partial information can make a difference. Indeed, numerous examples are known where quantum bits are more powerful than the same number of classical bits (see e.g.,~\cite{Nayak99,ANTV99,KoMaRe05,GKW:sepcc}). In this light, it is natural to study the potential  advantage offered by quantum information with respect to specific tasks.

We have taken a step in this direction by showing that certain schemes for the bounded storage model which are secure in the presence of classical adversaries are also secure in the presence of  adversaries who are in control of quantum storage. Surprisingly, the corresponding security parameters are almost the same 
for the quantum and the classical case when only a single bit is extracted. It is straightforward to extend and reformulate this result in terms of communication complexity. It then states that there can not be a large separation between the one-way average-case quantum and classical communication complexities of a boolean function. 

This is in sharp contrast to the case of extractors which output several bits. There are extractors that provide security in the classical bounded-storage model, but can not safely be used against quantum adversaries~\cite{GKW:sepcc}. Nevertheless, it is possible to give a family of constructions that yield secure bits; this is our main contribution.

While our extractors provide security against quantum adversaries, their parameters are far from optimal. Future work can focus on improving these constructions.

\section*{Acknowledgements}
RK thanks Ueli Maurer and Renato Renner for interesting discussions about bounded-storage cryptography. He would like to thank IBM Watson for their hospitality during the summer, and acknowledges
support by  the European Commission through the FP6-FET Integrated Project SCALA, CT-015714. BMT would like to thank Yevgeniy Dodis and Roberto Oliveira for many discussions on the security of the bounded-storage model. BMT acknowledges support by the NSA and the ARDA through ARO contract number W911NF-04-C-0098. 

The authors also thank Ronald de Wolf for helpful comments and for pointing out a mistake in the statement of Remark~\ref{rem:gavkempdewolf}, and thank Dodis for the suggestion to consider independent randomizers.

\appendix

\section{Properties of the nonuniformity\label{app:properties}}
We summarize a few properties of the non-uniformity  in this section.  

The non-uniformity $\nonun{P}{Z}{W}$ of $Z$ given $W$ can be viewed as the average distance of the conditional distribution $P_{Z|W=w}$ to the uniform distribution, for a random choice of $w\leftarrow P_W$, that is
\begin{align}
\nonun{P}{Z}{W}=\ExpE{w\leftarrow
P_W}[d(Z|W=w)]\label{eq:nonuniformityaverage}
\end{align}
 More generally,
for a ccq-state $\rho_{ZWQ}$, where $W$ and $Q$
are not necessarily independent, the non-uniformity of $Z$ given $WQ$ can be written as an average of the corresponding non-uniformities with respect to the conditional states $\rho_{ZQ|W=w}$. 
This is a direct consequence of Eq.~\eqref{eq:rhosigmadistance}. In formula, we have
\begin{align}\label{eq:nonuniformclassicalaverage}
\nonun{\rho}{Z}{WQ}=\ExpE{w\leftarrow P_W}
[\nonuncond{\rho}{Z}{Q}{W=w}]\ .
\end{align}
In particular, we can write
\begin{align}
\nonun{\rho}{\bigExt(X,Y)}{YQ}=\ExpE{y\leftarrow P_{\uniform_\cY}}
[\nonuncond{\rho}{\bigExt(X,y)}{Q}{Y=y}]\ ,\label{eq:expynonuniformity}
\end{align}
where the term in brackets is equal to $\nonun{P}{\bigExt(X,y)}{Q}$ when $Y$ and $Q$ are independent (which is usually the case in this paper).

Finally, we point out that conditioning on independent random variables leaves the non-uniformity invariant, that is
\begin{align}\label{eq:nonuniformindependent}
\nonun{\rho}{Z}{VQ}=\nonun{\rho}{Z}{Q}\ ,
\end{align}
if $\rho_{ZQV}=\rho_{ZQ}\otimes\rho_V$. This follows from Eq.~\eqref{eq:nonuniformclassicalaverage}.

\section{Proofs of Section \ref{sec:propextract}}
\label{app:a}

\begin{proof}[Proof of Proposition~\ref{lem:guessingprobability}]
Consider a random variable $\hat{X}$ defined by a channel $P_{\hat{X}|E}$
which takes a value $\hat{x}$ for which
$P_{X|E=e}(\hat{x})=2^{-H_\infty(X|E=e)}$ with certainty, for every $e\in\cE$. Clearly, we have
\begin{align*}
2^{-\hguessn{X}{E}}=\Pr[X=\hat{X}]=\ExpE{e\leftarrow P_E}
\bigl[2^{-H_\infty(X|E=e)}\bigr]\ .
\end{align*}
By Markov's inequality and Eq.~\eqref{eq:maximumsuccprob}, this implies
that 
\begin{align*}
\Pr_{e\leftarrow P_E}\bigl[H_\infty(X|E=e)\leq \hguessn{X}{E}-\log
1/\varepsilon \bigr]&\leq \epsilon\ .
\end{align*}
The result then follows by convexity, using the fact that
$\nonun{P}{\bigExt(X,Y)}{YE}=\ExpE{e\leftarrow P_E}
[\nonuncond{P}{\bigExt(X,Y)}{Y}{E=e}]$ because of Eq.~\eqref{eq:nonuniformityaverage}.
\end{proof}

Lemma~\ref{lem:wolflemmaclassical} can be seen as a special case of Lemma~\ref{lem:wolflemmaclassical}$^\prime$. Their proofs are analogous, but we include both here, as the classical proof is instructive for the quantum generalization.

\begin{proof}[Proof of Lemma~\ref{lem:wolflemmaclassical}]
Let $a_{v,w}:=2^{-\hguesscond{X}{E}{V=v,W=w}}$ for all
$(v,w)\in\cV\times \cW$. By definition,
\begin{align*}
a_{v,w}&=\sum_{e\in\cE} P_{E|V=v,W=w}(e)\max_{x\in\cX}
P_{X|E=e,V=v,W=w}(x)\ .
\end{align*}
In particular,
\begin{align*}
P_{VW}(v,w)a_{v,w}&=\sum_{e\in\cE} P_{EVW}(e,v,w)\max_{x\in\cX} P_{X|E=e,V=v,W=w}(x)\\
&=\sum_{e\in\cE}\max_{x\in\cX}P_{XEVW}(x,e,v,w)\\
\end{align*}
But by summing over $w\in\cW$
 \[ P_{XEVW}(x,e,v,w)\leq P_{XEV}(x,e,v)
\]
and thus
\begin{align*}
P_{VW}(v,w)a_{v,w}&\leq P_V(v)\sum_{e\in\cE}P_{E|V=v}(e)\max_{x\in\cX}P_{X|V=v,E=e}(x)\\
&=P_V(v)\sum_{e\in\cE}P_{E}(e)\max_{x\in\cX}P_{X|E=e}(x)\\
&=P_V(v)2^{-\hguessn{X}{E}}
\end{align*}
where we used the independence of $XE$ and $V$ in the second step
and the definition of $\hguessn{X}{E}$ to obtain the last identity.
We conclude that
\begin{align*}
\ExpE{(v,w)\leftarrow P_{VW}} [a_{v,w}]&\leq
2^{H_0(W)-\hguessn{X}{E}}
\end{align*}
It is easy to see that
\begin{align*}
\ExpE{(v,w)\leftarrow P_{VW}} [a_{v,w}]=2^{-\hguessn{X}{VWE}},
\end{align*}
which proves our first claim. We then use Markov's inequality to
obtain
\begin{align*}
\Pr_{(v,w)\leftarrow P_{VW}}\Bigl[a_{v,w}\geq \frac{1}{\varepsilon}
2^{H_0(W)-\hguessn{X}{E}}\Bigr]\leq \varepsilon\ ,
\end{align*}
which is our second claim.
\end{proof}

\begin{proof}[Proof of Lemma~\ref{lem:wolflemmaclassical}$^\prime$]
By assumption, $\rho_{XVWQ}$ has the form
\[
\rho_{XVWQ}=\sum_{(x,v,w)\in\cX\times\cV\times\cW}
P_{XVW}(xvw)\proj{xvw}\otimes\rho_x\ .
\] For every $(v,w)\in\cV\times \cW$, let $\cE^{v,w}:=\{E^{v,w}_x\}_{x\in\cX}$ be the {\rm POVM} which
maximizes the expression in the definition of
$\hguesscond{X}{Q}{V=v,W=w}$. We define the operators $\{F_x^v\}_{x\in\cX}$ by
\[
F_x^v:=2^{-H_0(W)}\sum_{w\in\cW} E^{v,w}_x\ .
\]
It is easy to see that $\cF^v:=\{F_x^v\}_{x\in\cX}$ forms a {\rm
POVM} for every $v\in\cV$, and the operator inequality
$E^{v,w}_x\leq 2^{H_0(W)} F_x^v$ holds. In particular,
\begin{align}\label{eq:trefineq}
\tr(E^{v,w}_x\rho_x)\leq 2^{H_0(W)}\tr(F^v_x\rho_x)
\end{align}
for all $(x,v,w)\in\cX\times\cV\times\cW$. Let us introduce the
abbreviation
\begin{align}\label{eq:avwdef}
a_{v,w}:=2^{-\hguesscond{X}{Q}{V=v,W=w}}\ .
\end{align}
for every $(v,w)\in\cV\times \cW$. By definition
and Eq.~\eqref{eq:trefineq},
\begin{align*}
a_{v,w}&=\sum_{x\in\cX} P_{X|V=v,W=w}(x)\tr(E^{v,w}_x\rho_x)\\
&\leq 2^{H_0(W)}\sum_{x\in\cX} P_{X|V=v,W=w}(x)\tr(F^v_x\rho_{x})\ .
\end{align*}
 We thus have
\begin{align}
\ExpE{(v,w)\leftarrow P_{VW}}[a_{v,w}] &\leq
2^{H_0(W)}\sum_{(x,v)\in\cX\times\cV}P_{XV}(x,v)\tr(F_x^v\rho_x)\nonumber\\
&=2^{H_0(W)}\ExpE{v\leftarrow P_V}\Bigl[ \sum_{x\in\cX}
P_{X|V=v}(x)\tr(F^v_x\rho_x) \Bigr]\ .\label{eq:hguessnxqinfirst}
\end{align}
But for every $v\in\cV$,
\begin{align}
\sum_{x\in\cX} P_{X|V=v}(x)\tr(F^v_x\rho_x)&=\sum_{x\in\cX} P_{X}(x)\tr(F^v_x\rho_x)\nonumber\\
&\leq 2^{-\hguessn{X}{Q}}\label{eq:hguessnxqin}
\end{align}
by the assumption that $\rho_{XV}=\rho_X\otimes\rho_V$, and the
definition of the latter quantity. Combining Eqs.~\eqref{eq:hguessnxqin}
with~\eqref{eq:hguessnxqinfirst} and~\eqref{eq:avwdef} gives
\[
2^{-\hguessn{X}{VWQ}}=\ExpE{(v,w)\leftarrow P_{VW}}[a_{v,w}] \leq
2^{H_0(W)-\hguessn{X}{Q}}\ ,
\]
and the first claim follows. The second claim follows from Markov's
inequality, as in the proof of Lemma~\ref{lem:wolflemmaclassical}.
\end{proof}
\vspace{0.2cm}

\section{Proof of Refinement Lemma}
\label{app:refine} In the proof of Theorem
\ref{thm:mainquantumbinary} we have used the fact that applying
classical post-processing after a measurement ${\cal F}$ does not
increase the non-uniformity. We state this as a lemma; the proof is
trivial and follows from the triangle inequality.
\begin{lemma}\label{lem:refinementlemma}
Let $P_{E|X}$ be a channel, and let
$\cF:=\{F_x\}_{x\in\cX}$ be a {\rm POVM} on $\cQ$. Define the
operators
\[
E_e:=\sum_{x\in\cX} P_{E|X=x}(e) F_x
\]
for every $e\in\cE$. Then $\cE:=\{E_e\}_{e\in\cE}$ is a {\rm POVM},
and for any cq-state $\rho_{ZQ}$,
\[
\nonun{\rho}{Z}{\cE(Q)}\leq \nonun{\rho}{Z}{\cF(Q)}\\ .
\]
\end{lemma}
\begin{proof}
It is trivial to check that $\cE$ is indeed a {\rm POVM}. By
definition,
\begin{align}
\nonun{\rho}{Z}{\cE(Q)}&=\|\rho_{Z\cE(Q)}-\rho_{\uniform_{\cZ}}\otimes\rho_{\cE(Q)}\|\nonumber\\
&=\frac{1}{2}\sum_{(z,e)\in\cZ\times\cE} \alpha_{z,e}\
,\label{eq:sumalphazzprime}
\end{align}
where
\[
\alpha_{z,e}:=\|\tr((\proj{z}\otimes
E_{e})\rho_{ZQ})-\frac{1}{|\cZ|}\tr(E_{e}\rho_Q)\|\ .
\]
By the definition of $E_{e}$ and the triangle inequality,
\[
\alpha_{z,e}\leq \sum_{x\in\cX} P_{E|X=x}(e) \|\tr((\proj{z}\otimes
F_x)\rho_{ZQ})-\frac{1}{|\cZ|}\tr(F_x\rho_Q)\|\ .
\]
Combining this with Eq.~\eqref{eq:sumalphazzprime} gives the claim.
\end{proof}


\end{document}